\documentclass[aps, pra, twocolumn, superscriptaddress, floatfix]{revtex4}
\usepackage{amsmath}
\usepackage{braket}
\usepackage[hidelinks]{hyperref}
\usepackage{graphicx}
\usepackage{tikz}

\newcommand{\Op}[1]{\ensuremath{\mathsf{\hat{#1}}}}
\newcommand{\tol}{\operatorname{tol}}
\newcommand{\PE}{\text{PE}}
\newcommand{\Abs}[1]{\left|#1\right|}
\newcommand{\Norm}[1]{\left\lVert#1\right\rVert}
\newcommand{\tr}{\mathrm{tr}}

\newcommand{\SU}{\ensuremath{\text{SU}}}

\newcommand{\Stanford}{Edward L. Ginzton Laboratory, Stanford University, Stanford, CA 94305, USA.}
\newcommand{\ARL}{U.S. Army Research Laboratory, Computational and Information Sciences Directorate, Adelphi, MD 20783, USA.}
\newcommand{\UMBC}{Department of Physics, University of Maryland Baltimore County, Baltimore, Maryland 21250, USA}

\begin{document}

\title{Local Gradient Optimization of Modular Entangling Sequences}

\author{A. A. Setser}
\affiliation{\UMBC}
\author{M. H. Goerz}
\affiliation{\Stanford}
\affiliation{\ARL}
\author{J. P. Kestner}
\affiliation{\UMBC}

\begin{abstract}
Implementation of logical entangling gates is an important step towards
realizing a quantum computer.
We use a gradient-based optimization approach to find single-qubit
rotations which can be interleaved between applications of a noisy nonlocal gate
to dramatically suppress arbitrary logical errors, while steering the evolution
operator towards the perfectly entangling subset of $\SU(4)$ gates.
The modularity of the approach allows for application to any two-qubit system,
regardless of the Hamiltonian or details of the experimental implementation.
This approach is effective for both quasi-static and time-dependent
1/$f^{\alpha}$ noise.
We also show how the fidelity of the final operation depends on both the
fidelity of the local rotations and the noise strength.
\end{abstract}

\maketitle

\section{Introduction}

Quantum computing requires both local (single-qubit) and nonlocal (entangling)
gate operations.
Fault-tolerant quantum computing requires that all operations on the physical
qubits can be performed with an error less than some ``quantum error correction
threshold'', the precise value depending on the particular choice of encoding
scheme.
Surface codes offer an attractively high threshold of roughly
1\%~\cite{WangPRA2011}, but even then it is desirable to reduce the physical
error rate further still in order to reduce the overhead associated with the
code.

Errors generally may result from leakage, i.e., loss of population out of the
logical subspace of a multilevel system, from \emph{quantum} noise, i.e.,
dissipation, or from \emph{classical} noise, i.e., fluctuations in the system
parameters or control fields.
Here, we will focus on (effective) spin-1/2 systems where leakage is not a
concern.
Dissipation generally depends inherently on the specific qubit implementation
and encoding, and can be minimized only by realizing gate operations on a time
scale that is short relative to the lifetime of the qubit.
Thus, we focus on reducing the sensitivity to classical noise.

In most physical systems, nonlocal gates have significantly higher errors than
local gates.
For single-qubit gates, noise can be countered using composite pulse sequences,
as has been experimentally demonstrated, for example, in
NMR~\cite{RevModPhys.76.1037}, trapped ions~\cite{PhysRevA.92.060301}, and spin
qubits~\cite{PhysRevLett.108.086802}.
For two-qubit entangling gates, however, the situation is more complicated.

One approach to two-qubit dynamical correction is pulse-shaping through optimal
control, varying several parameters in the system's Hamiltonian as a function of
time with numerically generated shaped pulses in order to minimize a target
functional.
Optimal control methods include both gradient-free methods, most commonly
Nelder-Mead~\cite{NelderCJ1965}, often possible through use of a low-dimensional
basis~\cite{CanevaPRA2011, RachPRA2015}; and gradient based methods,
gradient-ascent-pulse-engineering (GRAPE)~\cite{KhanejaJMR05,
deFouquieresJMR2011} and Krotov's method~\cite{KonnovARC99, SklarzPRA2002,
PalaoPRA2003, ReichJCP12}.
These methods have been shown to be a versatile tool for a wide range of tasks
in quantum engineering~\cite{GlaserEPJD2015, KochJPCM2016}.
This includes finding control fields that are robust with respect to
noise~\cite{ZhangPRA1994, KosutPRA2013, GoerzPRA2014, DongSR2015, HuangPRA2017}.
Although effective, this approach requires a precise and detailed knowledge of
both the control Hamiltonian and the form of the noise, including any
correlations between the two.

If one does not have such a complete model, an alternative approach is the
application of pulse sequences: using the relatively high fidelity of the local
gates one can construct a nonlocal gate sequence that cancels certain errors in
the elementary nonlocal
gates~\cite{PhysRevA.67.012317,PhysRevA.87.022323,PhysRevA.93.032340,Tomita2010}.
Knowledge of the system Hamiltonian is not required.
Here, single-qubit rotations are inserted between applications of an entangling
operation in such a way that \emph{any} systematic error present in the
entangling operation is canceled.
These sequences of single-qubit rotations can generally be applied repeatedly to
suppress systematic error to an arbitrary
level~\cite{PhysRevLett.98.180501,PhysRevLett.118.150502}.
However, those analytical results can require a large number of single-qubit
rotations, and accumulated imperfections in those rotations quickly diminish the
performance of the sequence.

In this paper, we present a new method which seeks to combine the efficiency of
numerical pulse-shaping with the agnosticism of composite pulse sequences.
We use a gradient-based optimization to numerically find a series of
single-qubit rotations to be interleaved between repeated application of a
slightly entangling gate such that the entire sequence performs a perfect
entangling gate while suppressing all possible logical errors.
The modularity of our approach allows for application to any system regardless
of the system's Hamiltonian or correlation of the noise with the control.
In addition, it performs well not only for systematic noise, but also for
time-dependent noise with a 1/f-type spectrum.

\section{Model and Optimization Method}

In order to counter the error in the numerical optimization, we sample an
ensemble of noisy system evolutions.
That is, we consider $M$ separate noise realizations, and require that the
optimized single-qubit operations perform well for the average over these
realizations~\cite{GoerzPRA2014}.

Within each noise realization, indexed by $m$, we assume that there exists a
total evolution operator such that $\ket{\psi(T)} =
\Op{U}^{(m)}\ket{\psi(t=0)}$, where $T$ is the total gate duration, and that
this operator can be broken into a series of $N$ steps.
The time evolution operator for the realization $(m)$ takes the form
\begin{equation} \label{U_full}
  \Op{U}^{(m)} =
    \prod_{n=N}^1
      \underbrace{\exp\left[-\frac{i\pi}{N}\Op{\sigma}_{ZZ}\right]}_{%
        \equiv \Op{Z}}\;
      \underbrace{\exp\left[-\frac{i}{N} \Op{\Delta}_{n}^{(m)} \right]}_{%
        \equiv \Op{D}_n^{(m)}}\;
      \Op{R}_n\,.
\end{equation}
with $n$ running backwards to account for time ordering.
Within each step of our evolution operator, $\Op{Z} \in \SU(4)$ with
$\Op{\sigma}_{ZZ} \equiv \Op{\sigma}_Z \otimes \Op{\sigma}_Z$ is a weakly
entangling gate.
Throughout this work $\Op{Z}$ is the $N$th root of a $2\pi$ phase gate.
In other words, the sequence consists of slicing an identity operation into $N$
equal pieces and inserting local rotations in between.
These local rotations steer the dynamics in arbitrary directions; the resulting
total gate $\Op{U}^{(m)}$ is not restricted to the diagonal gate that a pure
$\Op{\sigma}_{ZZ}$ interaction would induce.
We do not need to make any assumptions regarding how $\Op{Z}$ is implemented;
the effective Ising operation could itself be composed from any entangling
interaction plus local rotations.
However, the physical operation in the presence of noise consists of not only
the desired entangling gate $\Op{Z}$ but also a perturbation $\Op{D}_{n}^{(m)}$,
with
\begin{equation}
  \label{noise_operator}
  \Op{\Delta}^{(m)}_{n} \equiv \sum_{ij} \delta_{n,ij}^{(m)} \Op{\sigma}_{ij}
\end{equation}
where the $\delta_{n,ij}^{(m)}$ are the error coefficients and $\Op{\sigma}_{ij}
\equiv \Op{\sigma}_i \otimes \Op{\sigma}_j$.
The factor of $1/N$ in the exponential reflects our assumption that the error
scales with the size of the time slice.
Lastly, the $\Op{R}_n \in \SU(2) \otimes \SU(2)$ are controllable local
rotations that are used collectively to counter the effects of the noise
operators $\Op{D}_n^{(m)}$ while steering the total evolution $\Op{U}^{(m)}$ to
a perfect entangling gate.

We choose a parametrization in terms of Euler
angles~\cite{RoyalSocOpenSci.1.140145},
\begin{equation} \label{R}
  \begin{split}
    \Op{R}_n &=
      \exp\left[\frac{i\gamma_{1,n}}{2}\Op{\sigma}_Z\right]
      \exp\left[\frac{i\beta_{1,n}}{2}\Op{\sigma}_Y\right]
      \exp\left[\frac{i\alpha_{1,n}}{2}\Op{\sigma}_Z\right]
    \\ & \quad
      \otimes
      \exp\left[\frac{i\gamma_{2,n}}{2}\Op{\sigma}_Z\right]
      \exp\left[\frac{i\beta_{2,n}}{2}\Op{\sigma}_Y\right]
      \exp\left[\frac{i\alpha_{2,n}}{2}\Op{\sigma}_Z\right]\,.
  \end{split}
\end{equation}
The $6N$ parameters $\gamma_{1,1} \dots \alpha_{2,N}$ are the free control
parameters for the optimization.

We now consider the functional to be minimized by our numerical search method.
The functional should allow us to optimize towards two goals simultaneously:
\begin{enumerate}
  \item minimal sensitivity of the sequence to noise, and
  \item generation of a perfect entangler (PE), i.e., a gate which can produce
    a maximally entangled state from an unentangled
    one~\cite{PhysRevA.82.034301}.
\end{enumerate}
Reducing the sensitivity of Eq.~\eqref{U_full} to error $\Op{\Delta}^{(m)}$ can
be achieved by maximizing the trace overlap of $\Op{U}^{(m)}$ with the
equivalent operator that would have been generated in the absence of noise,
\begin{equation} \label{F}
  F(\Op{U}^{(m)}) =
    \frac{1}{16} \Abs{\tr\left(\Op{O}^{\dagger}\Op{U}^{(m)}\right)}^2\,,
    \quad
    \Op{O} = \prod_{n=N}^1 \Op{Z} \Op{R}_n\,,
\end{equation}
for every noise realization $(m)$.
The gate error for a particular noise realization corresponding to the fidelity
$F(\Op{U}^{(m)})$ is
\begin{equation}
  \label{gateerror}
  \epsilon(\Op{U}^{(m)}) = 1 - F(\Op{U}^{(m)})\,.
\end{equation}
Note that the target state includes the local rotations, and therefore changes
between every iteration of the optimization.
Optimization of Eq.~\eqref{F} serves to produce a known final gate which is
robust against noise.

However, even if we obtain the maximum value of $1$ for Eq.~\eqref{F} it does
not guarantee that the corresponding noise-free operation is an entangling one,
as the interleaved local rotations strongly impact the entanglement dynamics of
the final operation.
Hence, we must add another term to our functional to ensure the $\Op{U}$ is a
perfect entangler.

\begin{figure}[t]
  \includegraphics{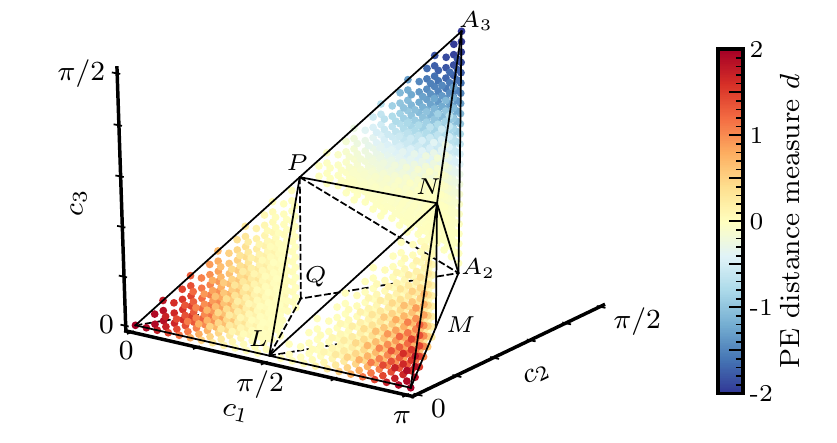}
  \caption{%
    The Weyl chamber, with the polyhedron of perfect entanglers (edges
    L--M--A$_2$--N--P--Q). The colored dots show the value of
    $d(c_1, c_2, c_3)$, Eq.~\eqref{d} numerically transformed from the Makhlin
    invariants $(g_1, g_2, g_3)$ to the Weyl chamber coordinates
    $(c_1, c_2, c_3)$.
    For visual clarity, the values for $d$ \emph{inside} the polyhedron are not
    shown -- they may be defined to zero.}%
  \label{weylchamber}
\end{figure}

For a gradient-based optimization scheme, the figure of merit for the
realization of a perfect entangler must analytically connect to the time
evolution operator $\Op{U}$.
It is not sufficient to simply calculate the concurrence of $\Op{U}$ as this is
a non-analytic quantity with an undefined gradient.
The key to optimizing for maximal entanglement is to note that any $\SU(4)$
matrix $\Op{U}$, can be written in the form of a Cartan decomposition $\Op{U} =
\Op{k}_1 \Op{A} \,\Op{k}_2$~\cite{KhanejaPRA2001,KrausPRA01,ZhangPRA2003}, with
the local operations $\Op{k}_1, \Op{k}_2 \in \SU(2) \otimes \SU(2)$, and
\begin{equation}
    \Op{A} =
    \exp\left[
      -\frac{i}{2}\left(
        c_1 \Op{\sigma}_{XX} +
        c_2 \Op{\sigma}_{YY} +
        c_3 \Op{\sigma}_{ZZ}\right)\right]
    \label{cartan_decomposition}
\end{equation}
representing a purely non-local operation $\in \SU(4)/(\SU(2) \otimes \SU(2))$.
The coefficients $(c_1, c_2, c_3)$ that parametrize $\Op{A}$ may be interpreted
as geometric coordinates that identify a two-qubit gate up to single-qubit
rotations.
The symmetries in Eq.~\eqref{cartan_decomposition} restrict $(c_1, c_2, c_3)$ to
form the Weyl chamber~\cite{ZhangPRA2003} shown in Fig.~\ref{weylchamber}.
All two-qubit gates which are perfect entanglers lie within a subset of the Weyl
chamber~\cite{WattsE2013}, the 7-faced polyhedron L--M--A$_2$--N--P--Q. The idea
of optimizing for maximal entanglement, developed in Refs.~\cite{WattsPRA2015,
GoerzPRA2015}, is to minimize the geometric distance to the closest surface of
the polyhedron.

The Weyl chamber coordinates $(c_1, c_2, c_3)$ still cannot be calculated
analytically from the gate $\Op{U}$, but the closely related Makhlin invariants
$(g_1, g_2, g_3)$ can~\cite{MakhlinQIP2002}. As there is a (non-analytic)
mapping $(g_1, g_2, g_3) \rightarrow (c_1, c_2, c_3)$~\cite{ZhangPRA2003}, the
desired distance can be expressed in terms of the Makhlin invariants
as~\cite{WattsPRA2015}
\begin{equation} \label{d}
  d(g_1, g_2, g_3) = g_3 \sqrt{g_1^2+g_2^2} - g_1\,.
\end{equation}
Converting Makhlin invariants back to Weyl chamber coordinates, the value of
$d(c_1, c_2, c_3)$ is indicated by color in Fig.~\ref{weylchamber}.

The value of $d$ in the top quadrant of the Weyl chamber (the so-called $W_1$
region, spanned by A$_2$--A$_3$--P--N) takes negative values.
In order to have a functional that can be minimized to achieve the objective, we
must thus change the sign.
To identify when a gate is in the $W_1$ region, we can calculate
\begin{equation}
  s=\pi-\cos^{-1}z_1-\cos^{-1}z_3\,.
\end{equation}
from the ordered roots ($z_1$, $z_2$, $z_3$) of the cubic
equation~\cite{ZhangPRA2003}
\begin{equation} \notag
  z^3-g_3z^2+(4\sqrt{g_1^2+g_2^2}-1)z+(g_3-4g_1)=0\,.
\end{equation}
Also, inside the polyhedron of perfect entanglers, $d$ takes values greater than
zero.
As we do not wish to bias the optimization towards perfect entanglers on the
surface of the polyhedron, we define the value of the functional for any perfect
entangler as zero.
Thus, in total the functional that should be minimized for the realization of a
perfect entangler is
\begin{equation} \label{D}
  \mathcal{D}(\Op{U})=
  \begin{cases}
  d  & \text{$d>0$ and $s>0$} \\
  -d & \text{$d<0$ and $s<0$} \\
  0  & \text{otherwise,}
  \end{cases}
\end{equation}
where $d = d(g_1(\Op{U}), g_2(\Op{U}), g_3(\Op{U}))$.
It yields zero if and only if $\Op{U}$ is a perfect entangler and is
positive otherwise~\cite{WattsPRA2015}.

The total optimization function is the sum of the gate error,
Eq.~\eqref{gateerror}, and the distance from the closest perfect entangler,
Eq.~\eqref{D}, averaged over all $M$ noise realizations,
\begin{equation} \label{J}
  J = \frac{1}{M} \sum_{m=1}^M \left(
    \epsilon(\Op{U}^{(m)}) + \mathcal{D}(\Op{U}^{(m)}) \right)\,.
\end{equation}
Minimization of the functional serves to produce a perfect entangler which is
robust against noise.

For the minimization of Eq.~\eqref{J} through variation of the $6N$ free
coefficients for the local error gates, we might in principle choose between
gradient-free and gradient-based methods.
While gradient-free methods are easy to implement, they show slow convergence
except for a small number of optimization parameters~\cite{EPJQT.2.21}.
Therefore, we look towards a gradient-based search method, which modifies every
control parameter $\eta_i \in \{\gamma_{1,1} \dots \alpha_{2,N}\}$ based on the
derivative of the functional, $\frac{\partial J}{\partial \eta_i}$.
The optimization is performed through the SciPy optimization
package~\cite{scipy}, using the L-BFGS-B
algorithm~\cite{SIAMJSciComput.16.1190}.
This algorithm offers an increase in convergence and stability through the
estimation of the Hessian (the matrix of second derivatives of the optimization
functional). In fact, SciPy's implementation of the L-BFGS-B algorithm also
allows to estimate the gradient of $J$ numerically.
This works for a moderate number of optimization parameters, and is an
attractive proposition in our case, as the analytical gradient of Eq.~\eqref{J}
is exceedingly tedious to calculate.
However, we note that Eq.~\eqref{J} was specifically written to guarantee the
existence of a well-defined analytical gradient.
For larger sequence lengths $N$ than we will consider here, the explicit
calculation of the gradient would eventually become necessary, but here we
proceed with the numerical approximation.

The convergence and success of the minimization for a single sequence length,
$N$, via the L-BFGS-B algorithm requires a reasonable choice of ``guess''
parameters as a starting point.
In order to maximize efficiency, if the greatest divisor of $N$ is $d$, we
repeat the solution for the length-$d$ sequence $\frac{N}{d}$ times to use as
the initial guess for the length-$N$ sequence.
This ensures that longer sequences will be constructed from shorter sequences
which have already been optimized to be robust against noise.
For prime-numbered sequence lengths, the local rotations are initialized as
identity operations.

When performing the optimization, we impose termination conditions by specifying
tolerances both for the functional and for the gradient of the functional.
The optimization terminates when either of the following two conditions is
fulfilled.
\begin{equation} \label{tolerance}
  \begin{aligned}
  \frac{J^k-J^{k+1}}{\max\{\Abs{J^k}, \Abs{J^{k+1}}, 1\}} & \leq \tol_J, \\
  \Norm{\text{proj}(\nabla{J})}_{\infty} & \leq \tol_{\nabla{J}}\,.
  \end{aligned}
\end{equation}
where $\text{proj}(\nabla{J})$ is the projection of the gradient vector onto the
space tangent to the active optimization bounds~\cite{ZhuATMS97}.
We find $\tol_J = \tol_{\nabla{J}} = 2.2 \times 10^{-6}$ to be a sufficient
value to identify the minimal error in the optimizations discussed in the
following section.

\begin{figure}[t]
  \includegraphics{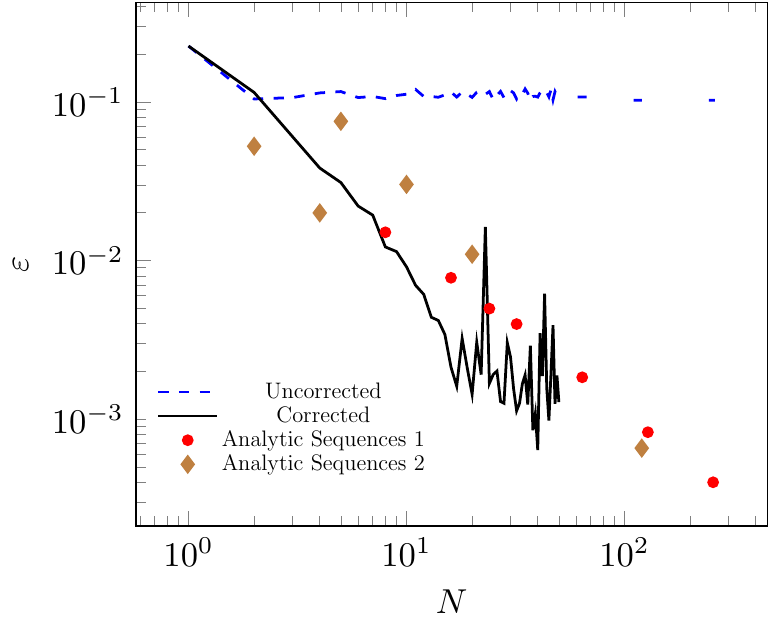}
  \caption{%
    Gate error in relation to sequence length for quasistatic nonlocal noise,
    assuming access to perfect single-qubit rotations; plotted on a log-log
    scale.
    Both lines shown are to be read at discrete values of $N$.
    The first set of analytic sequences are as generated in
    Ref.~\cite{PhysRevLett.98.180501} and the second set in
    Ref.~\cite{PhysRevLett.118.150502} (although, strictly speaking, only the
    last point, at $N=120$, corresponds to a sequence designed for the generic
    error being treated here).}%
  \label{errorFree}
\end{figure}

\section{Optimization Results}
For the analysis of the optimized sequences, we will consider the two desired
results of the optimization -- realization of a noise-free gate, and steering
into the perfect entangling polyhedron -- separately.
The success of the optimization routine in producing a gate $\Op{U}$ robust
against noise is evaluated through the gate error averaged over noise
realizations,
\begin{equation} \label{I}
  \varepsilon = \frac{1}{M} \sum_{m=1}^M \epsilon(\Op{U}^{(m)})\,.
\end{equation}
The fidelity of the gate $\Op{U}^{(m)}$ in a noise realization $(m)$ with
respect to the closest perfect entangler is evaluated as $F_{\PE}(\Op{U}) \in
[0, 1]$, defined as~\cite{WattsPRA2015}
\begin{equation} \label{Fpe}
  F_{\PE}(\Op{U}^{(m)})=
  \begin{cases}
  \cos^2\left(\frac{c_1+c_2-\frac{\pi}{2}}{4}\right) & c_1+c_2 \leq \frac{\pi}{2} \\
  \cos^2\left(\frac{c_2+c_3-\frac{\pi}{2}}{4}\right) & c_2+c_3 \geq \frac{\pi}{2} \\
  \cos^2\left(\frac{c_1-c_2-\frac{\pi}{2}}{4}\right) & c_1-c_2 \geq \frac{\pi}{2} \\
  1 & \text{otherwise,}
  \end{cases}
\end{equation}
where $(c_1, c_2, c_3)$ are the Weyl chamber coordinates of the gate
$\Op{U}^{(m)}$.
For evaluation purposes, this fidelity, or equivalently, the error
\begin{equation}\label{Epe}
\epsilon_{\PE}(\Op{U}^{(m)}) = 1 - F_{\PE}(\Op{U}^{(m)}),
\end{equation}
is a somewhat more direct measure than the (analytically differentiable)
geometric distance to the polyhedron of perfect entanglers in the Weyl chamber
that was used to steer the optimization, Eq.~\eqref{D}.
Again, the error is averaged over all noise realizations.
\begin{equation} \label{Ipe}
  \varepsilon_{\PE} = \frac{1}{M} \sum_{m=1}^M \epsilon_{\PE}(\Op{U}^{(m)}) \,.
\end{equation}
All optimized solutions can be accessed in an online repository~\footnote{\url{https://gist.github.com/goerz/3e04cca5a5dc1e9bcabd247065c24ebe}}.

\subsection{Quasistatic Gate Noise}

We first consider quasistatic noise; that is, noise that is constant on the
timescale of the operation.
Mathematically, this means that $\Op{\Delta}_n^{(m)}$ in Eq.~\eqref{U_full} is
the same for all time steps (independent of $n$), i.e., $\delta_{n,ij}^{(m)}
\rightarrow \delta_{ij}^{(m)}$ in Eq.~\eqref{noise_operator}.
The error coefficients $\delta_{ij}^{(m)}$ are drawn randomly from a normal
distribution with a standard deviation of $\sigma=0.13$.
The standard deviation is chosen such that when no local rotations are made to
suppress noise, the initial error, Eq.~\eqref{I}, is approximately $10\%$, a
realistic experimental situation~\cite{NPJQuantumInformation.3.3}.
If we only use one noise realization instead of averaging over many, the
optimization runs faster and gives higher fidelities, but clearly the optimal
parameters will be tailored to work especially well for that specific noise
realization and will not work well for an arbitrary realization.
This lack of generality persists even when averaging the optimization over 10
noise realizations.
However, we find that averaging over on the order of 100 noise realizations is
sufficient to ensure that our results are robust against a general noise
realization, i.e., the optimal parameters returned by running the optimization
over any random set of 100 noise realizations remain essentially the same.

Figure~\ref{errorFree} shows the results of this optimization.
The gate error shows a steady decrease with increasing $N$ up to around $N=16$,
after which the returns for increasing $N$ greatly diminish.
Note that for the longer sequences there is a lot of variability in the fidelity
due to sequences at prime values of $N$ having a significant disadvantage
compared to those at non-prime values where the optimization can be initialized
to a known good point using the results from shorter sequences.
While not shown, the values for $\varepsilon_{\PE}$ are less than or equal to
$10^{-8}$ for all sequences with $N \geq 2$ (including primes); with a majority
of sequence lengths resulting in values of $\varepsilon_{\PE}=0$.

For reference, we also show some points corresponding to analytical results
which use local rotations to achieve error
suppression~\cite{PhysRevLett.118.150502, PhysRevLett.98.180501}.
Ref.~\cite{PhysRevLett.98.180501} uses $N=8k$ applications, with $k$ an integer,
of an entangling Hamiltonian with local $\pi$ pulses about $x$ and $y$
interspersed to isolate a desired $\sigma_{ZZ}$ term in the Hamiltonian, with
residual error terms scaling like $1/N$.
In contrast, our present results show scaling closer to $1/N^2$ for small $N$.
Our results saturate around $N=16$, so eventually the analytical results do
better, but one has to go up all the way to $N=256$ to achieve similar
performance with the scheme of Ref.~\cite{PhysRevLett.98.180501}.
Ref.~\cite{PhysRevLett.118.150502} uses $N=120$ entangling operations with local
rotations (mostly $\pi$ rotations) to cancel the leading order error completely
generally.
For smaller initial noise, this approach is quite superior to suppression as
some power of $N$, but for the large noise values we have taken in this plot it
performs similar to the previous analytic scheme.
When some information is known about which type of errors are present, there are
shorter variants of with $N=$2, 4, 5, 10, and 20.
We show their performance as well, although they are not designed for the
general error case we are considering here.
Compared to Refs.~\cite{PhysRevLett.118.150502, PhysRevLett.98.180501}, our
results show an order of magnitude improvement in error suppression for
sequences under $50$ segments.
However, in contrast to our results, both analytical methods allow concatenation
to achieve, in principle, arbitrarily small errors at very large $N$.

So far, we have assumed that the local operations $\Op{R}_n$ that correct the
error can be implemented perfectly.
We now consider the effect of noise also in these local gates, i.e.,
imperfections in the control.
To this end, we randomly perturb each of the $6N$ Euler angles $\eta_i \in
\{\gamma_{1,1} \dots \alpha_{2,N}\}$ in Eq.~\eqref{R} according to
\begin{equation} \label{perturb}
  \eta_i \rightarrow \eta_i^{\prime} = \eta_i(1+\delta_{\eta})\,,
\end{equation}
where $\delta_{\eta}$ is an error coefficient drawn randomly from a normal
distribution with a standard deviation of $\sigma=0.01$.
This choice corresponds to the local rotations having experimentally realistic
fidelities of approximately $99.9\%$~\cite{Nature.508.500503}, calculated as
\begin{equation} \label{F_R}
  F_R =
    \frac{1}{16} \Abs{%
      \tr\left(
        \Op{R}^{\dagger}\left(
          \gamma^{\prime}_{1},\ldots,\alpha^{\prime}_{2}\right)
        \Op{R}\left(\gamma_{1},\ldots,\alpha_{2}\right)
      \right)}^2\,,
\end{equation}
and averaged over both $1000$ different sets of error coefficients and $1000$
sets of angles drawn randomly from a uniform distribution ranging from $-4\pi$
to $4\pi$.

\begin{figure}[t]
  \includegraphics{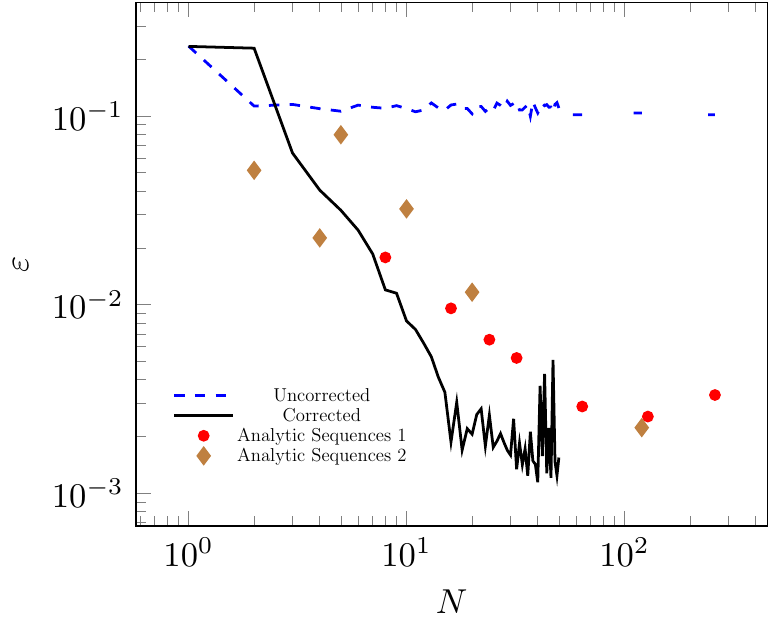}
  \caption{%
    Gate error in relation to sequence length for the noisy-rotation case,
    plotted on a log-log scale.
    All sources of error are quasistatic.
    The first set of analytic sequences are as generated in
    Ref.~\cite{PhysRevLett.98.180501} and the second set in
    Ref.~\cite{PhysRevLett.118.150502}}%
  \label{rotationErrors}
\end{figure}

The results of this optimization case are shown in Fig.~\ref{rotationErrors}.
The fidelity again scales favorably with $N$ compared to known analytic
sequences~\cite{PhysRevLett.118.150502, PhysRevLett.98.180501}, as shown in
Fig.~\ref{rotationErrors}.
The scaling does not continue for arbitrarily large sequence length, eventually
saturating, but now even the analytical results also saturate due to the
accumulation local noise.
In contrast, the performance of our sequences optimized in the presence of local
noise is only very slightly worse than the performance of the optimal sequences
in the absence of local noise.
The maximum fidelity achieved with these noisy local rotations is still
$99.90\%$, compared to $99.94\%$ fidelity achieved in the case of perfect local
rotations.
Thus, the effect of realistic errors in the single-qubit rotations only
marginally affects the optimal achievable performance.

Furthermore, the sequence optimized in the presence of local errors does not
sacrifice any ability to suppress nonlocal errors; we have checked that using
the new set of $6N$ parameters in the case of only nonlocal noise does not have
worse performance than the parameters obtained by optimizing specifically for
the case of nonlocal noise only.
The optimization places no restrictions on the rotation angles.
Without local noise, we find that angles remain in $[-\pi, \pi]$.
When local noise is introduced, however, the resulting angles are in the range
$[-4\pi, 4\pi]$, i.e., some rotations must involve multiple cycles around the
Bloch sphere in order for local errors to accumulate in an ultimately
self-negating way.
This is also seen in some analytic pulse
sequences~\cite{WangNatComm2012,KestnerPRL2013}.
These multiple cycles can be realized physically by, e.g., increasing the
rotation time.
As with the local-noise-free case, we have $\varepsilon_{\PE}\leq10^{-8}$ for
all sequences with $N\geq2$.

Recall that the optimization did not target a specific operation a priori, but
once the optimization is done once can characterize the specific robust perfect
entangler produced.
For instance, the operation produced by the $N=16$ optimization is, in Cartan
decomposed form,
\begin{align} \notag
  \begin{split}
    \Op{U}=\Op{k}_1\exp\{\frac{-i}{2}&\left(\right.2.250\Op{\sigma}_{XX}+0.809\Op{\sigma}_{YY}
    \\
    &+0.018\Op{\sigma}_{ZZ}\left.\right)\}\Op{k}_2\,.
  \end{split}
\end{align}
With the local operations $\Op{k}_1$ and $\Op{k}_2$ parametrized using Pauli
vectors,
\begin{widetext}
\begin{align} \notag
&\Op{k}_1=\exp\{-i( -0.827\Op{\sigma}_X+ 0.527\Op{\sigma}_Y-0.865\Op{\sigma}_Z)\} \otimes
\exp\{-i( -1.292\Op{\sigma}_X+ 0.006\Op{\sigma}_Y+1.589\Op{\sigma}_Z)\} \\ \notag
&\Op{k}_2=\exp\{-i( -0.993\Op{\sigma}_X +0.987\Op{\sigma}_Y-0.793\Op{\sigma}_Z)\} \otimes
\exp\{-i( 0.268\Op{\sigma}_X -0.965\Op{\sigma}_Y-1.769\Op{\sigma}_Z)\}.
\end{align}
\end{widetext}
Eq.~\eqref{Fpe} can be used to confirm that the produced operation is indeed a
perfect entangler.

\begin{figure}[t]\vspace*{-0.5cm}\includegraphics[scale=0.97]{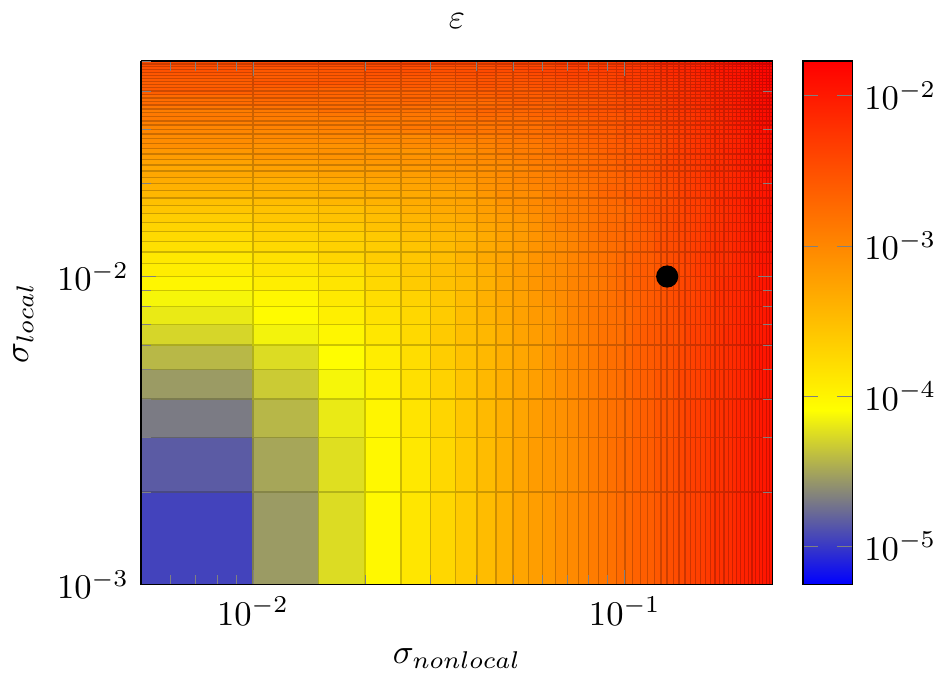}
  \caption{Gate error of the $N=16$ solution obtained in the presence of quasistatic
    noise in both the local and nonlocal operations, in relation to varying
    standard deviations of noise strengths. Plotted on a log-log-log scale.
    The solution obtained was optimized at $\sigma_{nonlocal}=0.13$,
    $\sigma_{local}=0.01$ (marked on plot), which corresponds to an uncorrected $90\%$
    fidelity in the nonlocal operations and a $99.9\%$ fidelity in the locals.}%
  \label{contour_plot}
\end{figure}

While the optimization was performed with an approximate range of values for the
noise strength, we wish to see how the solutions hold under varying noise
strengths.
Figure~\ref{contour_plot} shows the gate error for the $N=16$ solution generated
under quasistatic noise for a range of noise strengths, both in the local and
nonlocal operations.
Here we see that the obtained solution works well for all noise values which are
less than or equal to the values used in the optimization.
The noise strengths used in the optimization can be increased, which causes the
obtained solutions to hold for the new range of larger noise values.
However, the post-optimization fidelities at each $N$ will be smaller, due to
the difficulty of suppressing the now stronger noise.
For instance, if we instead optimize sequences with an initial $95\%$ fidelity
in the local operations and an $80\%$ fidelity in the nonlocals, the $N=16$
sequence would now hold for a range of noise values greater than what is shown
in Fig.
\ref{contour_plot}.
However, when this sequence is now evaluated at the (original) noise values
$\sigma_{nonlocal}=0.13, \sigma_{local}=0.01$, a fidelity of only $99.5\%$ is
obtained rather than the $99.9\%$ obtained under the previous optimization.
Thus, by choosing the noise strengths under which the optimization is performed,
one can trade between better results for a smaller range of noise values or more
modest gains that hold for a larger range of noise.

\subsection{Time-Dependent Gate Noise}

We now turn to noise that changes on a faster time scale than the gate operation
time.
In this case, each $\Op{\Delta}_n^{(m)}$ in Eq.~\eqref{U_full} is unique, but
taken to be correlated consistent with the properties of 1/$f^{\alpha}$
noise~\cite{ElectronicFluctuationsInSolids}.
The same is taken to be true for the noise in the local rotations, $\delta_n$ in
Eq.~\eqref{perturb}.
For concreteness, in the following we will take $\alpha=0.7$, consistent with
charge noise measurements in semiconductor spin
qubits~\cite{PhysRevLett.110.146804}.
The time-dependent noise is constructed by first generating a signal that
randomly switches between $\pm 1$; also known as a random telegraph process.
The switching events themselves have a Poisson distribution.
A single process has a characteristic time constant $\tau$ which determines the
average time spent in one state over the course of the many switches, with a
relaxation rate of $\nu=\frac{1}{2\tau}$.
The power spectral density of the random telegraph signal is a Lorentzian of
width $\nu$.
The 1/$f^{\alpha}$ noise is then generated approximately by superimposing a
finite number of fluctuators with relaxation rates evenly spaced logarithmically
between $\nu_{\min}$ and $\nu_{\max}$.

\begin{figure}[t]
  \includegraphics{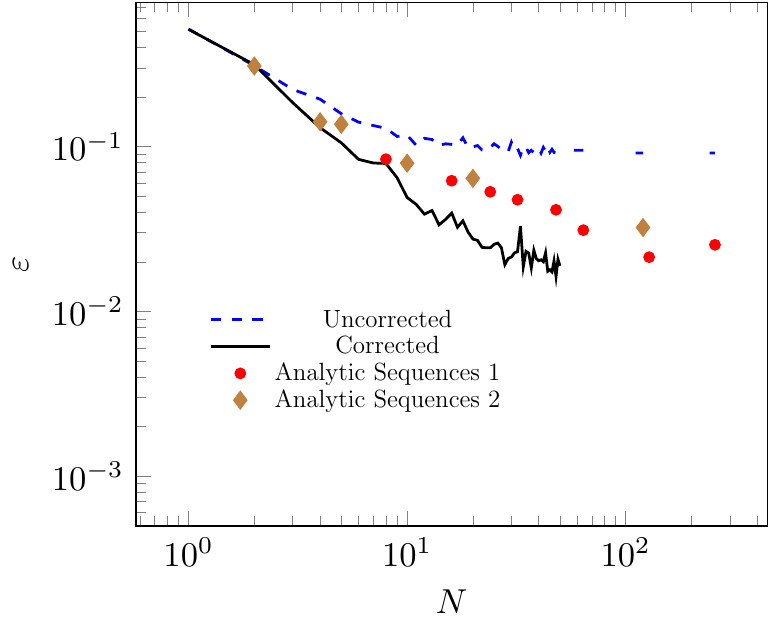}
  \caption{%
    Gate error in relation to sequence length, plotted on a log-log scale,
    assuming that both the local and nonlocal noise properties are consistent
    with those of 1/$f$ noise.}%
  \label{oneOverF}
\end{figure}

We assign errors to the various segments of the pulse in the following way.
Considering an arbitrary entangling gate time of $T$, we take $\nu_{\min} =
\frac{1}{2(10 T)}$ and $\nu_{\max} = \frac{1}{2(T/10)}$ and we use ten random
telegraph fluctuators to generate the noise.
For a single channel (fixed $ij$), the error coefficients for separate segments
$n$ are drawn by sampling a given time trace of the $1/f$ noise at different
times $t_n$, where $t_n$ is chosen randomly (via a uniform distribution) within
the interval $\left[(n-1)\frac{T}{N}, n\frac{T}{N}\right]$ causing the noise
within each channel to vary over time.
Noise in the local operations is handled in the same way.

As in the quasistatic noise case, we scale the amplitude of the random telegraph
signals such that the local rotations possess an approximate $99.9\%$ fidelity
on average and the uncorrected entangling operation has a fidelity of roughly
$90.0\%$.
In this case, the standard deviations of the sets of discretely sampled $1/f$
noise making up the local and nonlocal errors are $\sigma_{\text{local}}=0.006$
and $\sigma_{\text{nonlocal}}=0.2$.

The performance of the optimization routine in the time-dependent noise case is
shown in Figure~\ref{oneOverF}.
Similar to the quasistatic noise case, the value of $\varepsilon_{\PE}$ is less
than or equal to $10^{-8}$ for all sequences of length $N \geq 2$, with a
majority possessing $\varepsilon_{\PE}=0$.
Note that, unlike in the quasistatic case, the uncorrected error diminishes
somewhat as $N$ increases.
This reflects the fact that the time-dependent noise is less correlated than the
quasistatic noise and for long sequences a little of the noise can average out
even without any intervention.
Comparing gate fidelities, the optimization routine is approximately an order of
magnitude worse in the time-dependent case compared to the quasistatic case
($99.0\%$ fidelity compared to $99.9\%$), but still offers an order of magnitude
improvement compared to the uncorrected gate.
It should also be noted that rotations obtained in the presence of 1/$f$ noise
are also able to suppress quasistatic errors to the same levels that they were
suppressed in the 1/$f$ case, while the converse is not true.

In the case where there is $1/f$ nonlocal noise but
no local noise (as might be expected if the local gates were practically
instantaneous), we achieve a gate error of $5\times 10^{-3}$ at $N=30$, compared
to an error of $4\times 10^{-2}$ using the sequence of
Ref.~\cite{PhysRevLett.118.150502} at $N=120$ or $2\times 10^{-2}$ at length
$N=256$ using the sequence of Ref.~\cite{PhysRevLett.98.180501}.
Thus, for 1/$f$ noise as in the case of quasistatic noise, this optimization
routine has significantly better performance than known analytic pulse
sequences.

\section{Conclusion}
Under reasonable conditions, we have shown that primitive entangling gates of
fidelity near $90\%$ can be used to construct composite entangling gates with
error rates well below the $\sim 1\%$ fault tolerance threshold of the surface
code.
A possible difficulty lies in the fact that the interwoven optimized local
operations are not simple rational multiples of $\pi$, but consist of generic
angles.
If we restrict the optimization to select the rotations from a small set of
fractions of $\pi$, the optimization is no longer effective.
Thus, experimentally calibrating a set of 10-20 unique local rotations is the
cost one has to pay for boosting the fidelity of the nonlocal operation.

We have presented results from numerically optimizing a modular, composite pulse
sequence, requiring no knowledge of the underlying Hamiltonian.
The high-fidelity local rotations suppress arbitrary logical error in the
nonlocal operation, while steering the overall operation into the class of
perfect entanglers.
Our results show significant improvement over known analytical results,
requiring fewer local operations to achieve logical error suppression.
The modularity of this approach allows for application to any two-qubit system
regardless of the Hamiltonian.
The approach has been shown to provide significant error suppression in both
quasistatic and 1/$f^{0.7}$ noise cases.

\section*{Acknowledgments}
AAS and JPK acknowledge support by the National Science Foundation under Grant
No.~1620740.
MHG acknowledges support by ASD(R\&E) under their Quantum Science and
Engineering Program (QSEP), by the Army High Performance Computing Research
Center (AHPCRC) under contract Number W911NF-07-2-0027, and by the U.S. Army
Research Laboratory under Cooperative Agreement Number W911NF-16-2-0147.

\bibliography{modularEntanglingSequencesBib}

\begin{thebibliography}{45}
\expandafter\ifx\csname natexlab\endcsname\relax\def\natexlab#1{#1}\fi
\expandafter\ifx\csname bibnamefont\endcsname\relax
  \def\bibnamefont#1{#1}\fi
\expandafter\ifx\csname bibfnamefont\endcsname\relax
  \def\bibfnamefont#1{#1}\fi
\expandafter\ifx\csname citenamefont\endcsname\relax
  \def\citenamefont#1{#1}\fi
\expandafter\ifx\csname url\endcsname\relax
  \def\url#1{\texttt{#1}}\fi
\expandafter\ifx\csname urlprefix\endcsname\relax\def\urlprefix{URL }\fi
\providecommand{\bibinfo}[2]{#2}
\providecommand{\eprint}[2][]{\url{#2}}

\bibitem[{\citenamefont{Wang et~al.}(2011)\citenamefont{Wang, Fowler, and
  Hollenberg}}]{WangPRA2011}
\bibinfo{author}{\bibfnamefont{D.~S.} \bibnamefont{Wang}},
  \bibinfo{author}{\bibfnamefont{A.~G.} \bibnamefont{Fowler}},
  \bibnamefont{and} \bibinfo{author}{\bibfnamefont{L.~C.~L.}
  \bibnamefont{Hollenberg}}, \bibinfo{journal}{Phys. Rev. A}
  \textbf{\bibinfo{volume}{83}}, \bibinfo{pages}{020302}
  (\bibinfo{year}{2011}).

\bibitem[{\citenamefont{Vandersypen and Chuang}(2005)}]{RevModPhys.76.1037}
\bibinfo{author}{\bibfnamefont{L.~M.~K.} \bibnamefont{Vandersypen}}
  \bibnamefont{and} \bibinfo{author}{\bibfnamefont{I.~L.}
  \bibnamefont{Chuang}},
  \bibinfo{journal}{\href{https://link.aps.org/doi/10.1103/RevModPhys.76.1037}{Rev.
  Mod. Phys.}} \textbf{\bibinfo{volume}{76}}, \bibinfo{pages}{1037}
  (\bibinfo{year}{2005}).

\bibitem[{\citenamefont{Mount et~al.}(2015)\citenamefont{Mount, Kabytayev,
  Crain, Harper, Baek, Vrijsen, Flammia, Brown, Maunz, and
  Kim}}]{PhysRevA.92.060301}
\bibinfo{author}{\bibfnamefont{E.}~\bibnamefont{Mount}},
  \bibinfo{author}{\bibfnamefont{C.}~\bibnamefont{Kabytayev}},
  \bibinfo{author}{\bibfnamefont{S.}~\bibnamefont{Crain}},
  \bibinfo{author}{\bibfnamefont{R.}~\bibnamefont{Harper}},
  \bibinfo{author}{\bibfnamefont{S.-Y.} \bibnamefont{Baek}},
  \bibinfo{author}{\bibfnamefont{G.}~\bibnamefont{Vrijsen}},
  \bibinfo{author}{\bibfnamefont{S.~T.} \bibnamefont{Flammia}},
  \bibinfo{author}{\bibfnamefont{K.~R.} \bibnamefont{Brown}},
  \bibinfo{author}{\bibfnamefont{P.}~\bibnamefont{Maunz}}, \bibnamefont{and}
  \bibinfo{author}{\bibfnamefont{J.}~\bibnamefont{Kim}},
  \bibinfo{journal}{\href{https://link.aps.org/doi/10.1103/PhysRevA.92.060301}{Phys.
  Rev. A}} \textbf{\bibinfo{volume}{92}}, \bibinfo{pages}{060301}
  (\bibinfo{year}{2015}).

\bibitem[{\citenamefont{Medford et~al.}(2012)\citenamefont{Medford,
  Cywi\ifmmode~\acute{n}\else \'{n}\fi{}ski, Barthel, Marcus, Hanson, and
  Gossard}}]{PhysRevLett.108.086802}
\bibinfo{author}{\bibfnamefont{J.}~\bibnamefont{Medford}},
  \bibinfo{author}{\bibfnamefont{L.}~\bibnamefont{Cywi\ifmmode~\acute{n}\else
  \'{n}\fi{}ski}}, \bibinfo{author}{\bibfnamefont{C.}~\bibnamefont{Barthel}},
  \bibinfo{author}{\bibfnamefont{C.~M.} \bibnamefont{Marcus}},
  \bibinfo{author}{\bibfnamefont{M.~P.} \bibnamefont{Hanson}},
  \bibnamefont{and} \bibinfo{author}{\bibfnamefont{A.~C.}
  \bibnamefont{Gossard}},
  \bibinfo{journal}{\href{https://link.aps.org/doi/10.1103/PhysRevLett.108.086802}{Phys.
  Rev. Lett.}} \textbf{\bibinfo{volume}{108}}, \bibinfo{pages}{086802}
  (\bibinfo{year}{2012}).

\bibitem[{\citenamefont{Nelder and Mead}(1965)}]{NelderCJ1965}
\bibinfo{author}{\bibfnamefont{J.~A.} \bibnamefont{Nelder}} \bibnamefont{and}
  \bibinfo{author}{\bibfnamefont{R.}~\bibnamefont{Mead}},
  \bibinfo{journal}{Comput. J.} \textbf{\bibinfo{volume}{7}},
  \bibinfo{pages}{308} (\bibinfo{year}{1965}).

\bibitem[{\citenamefont{Caneva et~al.}(2011)\citenamefont{Caneva, Calarco, and
  Montangero}}]{CanevaPRA2011}
\bibinfo{author}{\bibfnamefont{T.}~\bibnamefont{Caneva}},
  \bibinfo{author}{\bibfnamefont{T.}~\bibnamefont{Calarco}}, \bibnamefont{and}
  \bibinfo{author}{\bibfnamefont{S.}~\bibnamefont{Montangero}},
  \bibinfo{journal}{Phys. Rev. A} \textbf{\bibinfo{volume}{84}},
  \bibinfo{pages}{022326} (\bibinfo{year}{2011}).

\bibitem[{\citenamefont{Rach et~al.}(2015)\citenamefont{Rach, M\"uller,
  Calarco, and Montangero}}]{RachPRA2015}
\bibinfo{author}{\bibfnamefont{N.}~\bibnamefont{Rach}},
  \bibinfo{author}{\bibfnamefont{M.~M.} \bibnamefont{M\"uller}},
  \bibinfo{author}{\bibfnamefont{T.}~\bibnamefont{Calarco}}, \bibnamefont{and}
  \bibinfo{author}{\bibfnamefont{S.}~\bibnamefont{Montangero}},
  \bibinfo{journal}{Phys. Rev. A} \textbf{\bibinfo{volume}{92}},
  \bibinfo{pages}{062343} (\bibinfo{year}{2015}).

\bibitem[{\citenamefont{Khaneja et~al.}(2005)\citenamefont{Khaneja, Reiss,
  Kehlet, Schulte-Herbr{\"u}ggen, and Glaser}}]{KhanejaJMR05}
\bibinfo{author}{\bibfnamefont{N.}~\bibnamefont{Khaneja}},
  \bibinfo{author}{\bibfnamefont{T.}~\bibnamefont{Reiss}},
  \bibinfo{author}{\bibfnamefont{C.}~\bibnamefont{Kehlet}},
  \bibinfo{author}{\bibfnamefont{T.}~\bibnamefont{Schulte-Herbr{\"u}ggen}},
  \bibnamefont{and} \bibinfo{author}{\bibfnamefont{S.~J.}
  \bibnamefont{Glaser}}, \bibinfo{journal}{J. Magnet. Res.}
  \textbf{\bibinfo{volume}{172}}, \bibinfo{pages}{296} (\bibinfo{year}{2005}).

\bibitem[{\citenamefont{de~Fouquieres et~al.}(2011)\citenamefont{de~Fouquieres,
  Schirmer, Glaser, and Kuprov}}]{deFouquieresJMR2011}
\bibinfo{author}{\bibfnamefont{P.}~\bibnamefont{de~Fouquieres}},
  \bibinfo{author}{\bibfnamefont{S.}~\bibnamefont{Schirmer}},
  \bibinfo{author}{\bibfnamefont{S.}~\bibnamefont{Glaser}}, \bibnamefont{and}
  \bibinfo{author}{\bibfnamefont{I.}~\bibnamefont{Kuprov}},
  \bibinfo{journal}{\href{http://www.sciencedirect.com/science/article/pii/S1090780711002552}{JMR}}
  \textbf{\bibinfo{volume}{212}}, \bibinfo{pages}{412} (\bibinfo{year}{2011}).

\bibitem[{\citenamefont{Konnov and Krotov}(1999)}]{KonnovARC99}
\bibinfo{author}{\bibfnamefont{A.}~\bibnamefont{Konnov}} \bibnamefont{and}
  \bibinfo{author}{\bibfnamefont{V.~F.} \bibnamefont{Krotov}},
  \bibinfo{journal}{Autom. Rem. Contr.} \textbf{\bibinfo{volume}{60}}
  (\bibinfo{year}{1999}).

\bibitem[{\citenamefont{Sklarz and Tannor}(2002)}]{SklarzPRA2002}
\bibinfo{author}{\bibfnamefont{S.~E.} \bibnamefont{Sklarz}} \bibnamefont{and}
  \bibinfo{author}{\bibfnamefont{D.~J.} \bibnamefont{Tannor}},
  \bibinfo{journal}{Phys. Rev. A} \textbf{\bibinfo{volume}{66}},
  \bibinfo{pages}{053619} (\bibinfo{year}{2002}).

\bibitem[{\citenamefont{Palao and Kosloff}(2003)}]{PalaoPRA2003}
\bibinfo{author}{\bibfnamefont{J.~P.} \bibnamefont{Palao}} \bibnamefont{and}
  \bibinfo{author}{\bibfnamefont{R.}~\bibnamefont{Kosloff}},
  \bibinfo{journal}{Phys. Rev. A} \textbf{\bibinfo{volume}{68}},
  \bibinfo{pages}{062308} (\bibinfo{year}{2003}).

\bibitem[{\citenamefont{Reich et~al.}(2012)\citenamefont{Reich, Ndong, and
  Koch}}]{ReichJCP12}
\bibinfo{author}{\bibfnamefont{D.~M.} \bibnamefont{Reich}},
  \bibinfo{author}{\bibfnamefont{M.}~\bibnamefont{Ndong}}, \bibnamefont{and}
  \bibinfo{author}{\bibfnamefont{C.~P.} \bibnamefont{Koch}},
  \bibinfo{journal}{J. Chem. Phys.} \textbf{\bibinfo{volume}{136}},
  \bibinfo{pages}{104103} (\bibinfo{year}{2012}).

\bibitem[{\citenamefont{Glaser et~al.}(2015)\citenamefont{Glaser, Boscain,
  Calarco, Koch, K\"ockenberger, Kosloff, Kuprov, Luy, Schirmer,
  Schulte-Herbr\"uggen et~al.}}]{GlaserEPJD2015}
\bibinfo{author}{\bibfnamefont{S.~J.} \bibnamefont{Glaser}},
  \bibinfo{author}{\bibfnamefont{U.}~\bibnamefont{Boscain}},
  \bibinfo{author}{\bibfnamefont{T.}~\bibnamefont{Calarco}},
  \bibinfo{author}{\bibfnamefont{C.~P.} \bibnamefont{Koch}},
  \bibinfo{author}{\bibfnamefont{W.}~\bibnamefont{K\"ockenberger}},
  \bibinfo{author}{\bibfnamefont{R.}~\bibnamefont{Kosloff}},
  \bibinfo{author}{\bibfnamefont{I.}~\bibnamefont{Kuprov}},
  \bibinfo{author}{\bibfnamefont{B.}~\bibnamefont{Luy}},
  \bibinfo{author}{\bibfnamefont{S.}~\bibnamefont{Schirmer}},
  \bibinfo{author}{\bibfnamefont{T.}~\bibnamefont{Schulte-Herbr\"uggen}},
  \bibnamefont{et~al.}, \bibinfo{journal}{Eur. Phys. J. D}
  \textbf{\bibinfo{volume}{69}}, \bibinfo{pages}{279} (\bibinfo{year}{2015}).

\bibitem[{\citenamefont{Koch}(2016)}]{KochJPCM2016}
\bibinfo{author}{\bibfnamefont{C.~P.} \bibnamefont{Koch}}, \bibinfo{journal}{J.
  Phys.: Condens. Matter} \textbf{\bibinfo{volume}{28}},
  \bibinfo{pages}{213001} (\bibinfo{year}{2016}).

\bibitem[{\citenamefont{Zhang and Rabitz}(1994)}]{ZhangPRA1994}
\bibinfo{author}{\bibfnamefont{H.}~\bibnamefont{Zhang}} \bibnamefont{and}
  \bibinfo{author}{\bibfnamefont{H.}~\bibnamefont{Rabitz}},
  \bibinfo{journal}{Phys. Rev. A} \textbf{\bibinfo{volume}{49}},
  \bibinfo{pages}{2241} (\bibinfo{year}{1994}).

\bibitem[{\citenamefont{Kosut et~al.}(2013)\citenamefont{Kosut, Grace, and
  Brif}}]{KosutPRA2013}
\bibinfo{author}{\bibfnamefont{R.~L.} \bibnamefont{Kosut}},
  \bibinfo{author}{\bibfnamefont{M.~D.} \bibnamefont{Grace}}, \bibnamefont{and}
  \bibinfo{author}{\bibfnamefont{C.}~\bibnamefont{Brif}},
  \bibinfo{journal}{Phys. Rev. A} \textbf{\bibinfo{volume}{88}},
  \bibinfo{pages}{052326} (\bibinfo{year}{2013}).

\bibitem[{\citenamefont{Goerz et~al.}(2014)\citenamefont{Goerz, Halperin,
  Aytac, Koch, and Whaley}}]{GoerzPRA2014}
\bibinfo{author}{\bibfnamefont{M.~H.} \bibnamefont{Goerz}},
  \bibinfo{author}{\bibfnamefont{E.~J.} \bibnamefont{Halperin}},
  \bibinfo{author}{\bibfnamefont{J.~M.} \bibnamefont{Aytac}},
  \bibinfo{author}{\bibfnamefont{C.~P.} \bibnamefont{Koch}}, \bibnamefont{and}
  \bibinfo{author}{\bibfnamefont{K.~B.} \bibnamefont{Whaley}},
  \bibinfo{journal}{Phys. Rev. A} \textbf{\bibinfo{volume}{90}},
  \bibinfo{pages}{032329} (\bibinfo{year}{2014}).

\bibitem[{\citenamefont{Dong et~al.}(2015)\citenamefont{Dong, Chen, Qi,
  Petersen, and Nori}}]{DongSR2015}
\bibinfo{author}{\bibfnamefont{D.}~\bibnamefont{Dong}},
  \bibinfo{author}{\bibfnamefont{C.}~\bibnamefont{Chen}},
  \bibinfo{author}{\bibfnamefont{B.}~\bibnamefont{Qi}},
  \bibinfo{author}{\bibfnamefont{I.~R.} \bibnamefont{Petersen}},
  \bibnamefont{and} \bibinfo{author}{\bibfnamefont{F.}~\bibnamefont{Nori}},
  \bibinfo{journal}{Sci. Rep.} \textbf{\bibinfo{volume}{5}},
  \bibinfo{pages}{7873} (\bibinfo{year}{2015}).

\bibitem[{\citenamefont{Huang and Goan}(2017)}]{HuangPRA2017}
\bibinfo{author}{\bibfnamefont{C.-H.} \bibnamefont{Huang}} \bibnamefont{and}
  \bibinfo{author}{\bibfnamefont{H.-S.} \bibnamefont{Goan}},
  \bibinfo{journal}{\href{https://link.aps.org/doi/10.1103/PhysRevA.95.062325}{Phys.
  Rev. A}} \textbf{\bibinfo{volume}{95}}, \bibinfo{pages}{062325}
  (\bibinfo{year}{2017}).

\bibitem[{\citenamefont{Jones}(2003)}]{PhysRevA.67.012317}
\bibinfo{author}{\bibfnamefont{J.~A.} \bibnamefont{Jones}},
  \bibinfo{journal}{\href{https://link.aps.org/doi/10.1103/PhysRevA.67.012317}{Phys.
  Rev. A}} \textbf{\bibinfo{volume}{67}}, \bibinfo{pages}{012317}
  (\bibinfo{year}{2003}).

\bibitem[{\citenamefont{Ichikawa et~al.}(2013)\citenamefont{Ichikawa,
  G\"ung\"ord\"u, Bando, Kondo, and Nakahara}}]{PhysRevA.87.022323}
\bibinfo{author}{\bibfnamefont{T.}~\bibnamefont{Ichikawa}},
  \bibinfo{author}{\bibfnamefont{U.}~\bibnamefont{G\"ung\"ord\"u}},
  \bibinfo{author}{\bibfnamefont{M.}~\bibnamefont{Bando}},
  \bibinfo{author}{\bibfnamefont{Y.}~\bibnamefont{Kondo}}, \bibnamefont{and}
  \bibinfo{author}{\bibfnamefont{M.}~\bibnamefont{Nakahara}},
  \bibinfo{journal}{\href{https://link.aps.org/doi/10.1103/PhysRevA.87.022323}{Phys.
  Rev. A}} \textbf{\bibinfo{volume}{87}}, \bibinfo{pages}{022323}
  (\bibinfo{year}{2013}).

\bibitem[{\citenamefont{Cohen et~al.}(2016)\citenamefont{Cohen, Rotem, and
  Retzker}}]{PhysRevA.93.032340}
\bibinfo{author}{\bibfnamefont{I.}~\bibnamefont{Cohen}},
  \bibinfo{author}{\bibfnamefont{A.}~\bibnamefont{Rotem}}, \bibnamefont{and}
  \bibinfo{author}{\bibfnamefont{A.}~\bibnamefont{Retzker}},
  \bibinfo{journal}{\href{https://link.aps.org/doi/10.1103/PhysRevA.93.032340}{Phys.
  Rev. A}} \textbf{\bibinfo{volume}{93}}, \bibinfo{pages}{032340}
  (\bibinfo{year}{2016}).

\bibitem[{\citenamefont{Tomita et~al.}(2010)\citenamefont{Tomita, Merrill, and
  Brown}}]{Tomita2010}
\bibinfo{author}{\bibfnamefont{Y.}~\bibnamefont{Tomita}},
  \bibinfo{author}{\bibfnamefont{J.~T.} \bibnamefont{Merrill}},
  \bibnamefont{and} \bibinfo{author}{\bibfnamefont{K.~R.} \bibnamefont{Brown}},
  \bibinfo{journal}{\href{http://stacks.iop.org/1367-2630/12/i=1/a=015002}{New
  Journal of Physics}} \textbf{\bibinfo{volume}{12}}, \bibinfo{pages}{015002}
  (\bibinfo{year}{2010}).

\bibitem[{\citenamefont{Hill}(2007)}]{PhysRevLett.98.180501}
\bibinfo{author}{\bibfnamefont{C.~D.} \bibnamefont{Hill}},
  \bibinfo{journal}{\href{https://link.aps.org/doi/10.1103/PhysRevLett.98.180501}{Phys.
  Rev. Lett.}} \textbf{\bibinfo{volume}{98}}, \bibinfo{pages}{180501}
  (\bibinfo{year}{2007}).

\bibitem[{\citenamefont{Calderon-Vargas and
  Kestner}(2017)}]{PhysRevLett.118.150502}
\bibinfo{author}{\bibfnamefont{F.~A.} \bibnamefont{Calderon-Vargas}}
  \bibnamefont{and} \bibinfo{author}{\bibfnamefont{J.~P.}
  \bibnamefont{Kestner}},
  \bibinfo{journal}{\href{https://link.aps.org/doi/10.1103/PhysRevLett.118.150502}{Phys.
  Rev. Lett.}} \textbf{\bibinfo{volume}{118}}, \bibinfo{pages}{150502}
  (\bibinfo{year}{2017}).

\bibitem[{\citenamefont{Hamada}(2014)}]{RoyalSocOpenSci.1.140145}
\bibinfo{author}{\bibfnamefont{M.}~\bibnamefont{Hamada}},
  \bibinfo{journal}{\href{http://doi.org/10.1098/rsos.140145}{Royal Soc. Open
  Sci.}} \textbf{\bibinfo{volume}{1}}, \bibinfo{pages}{140145}
  (\bibinfo{year}{2014}).

\bibitem[{\citenamefont{Balakrishnan and
  Sankaranarayanan}(2010)}]{PhysRevA.82.034301}
\bibinfo{author}{\bibfnamefont{S.}~\bibnamefont{Balakrishnan}}
  \bibnamefont{and}
  \bibinfo{author}{\bibfnamefont{R.}~\bibnamefont{Sankaranarayanan}},
  \bibinfo{journal}{\href{https://link.aps.org/doi/10.1103/PhysRevA.82.034301}{Phys.
  Rev. A}} \textbf{\bibinfo{volume}{82}}, \bibinfo{pages}{034301}
  (\bibinfo{year}{2010}).

\bibitem[{\citenamefont{Khaneja et~al.}(2001)\citenamefont{Khaneja, Brockett,
  and Glaser}}]{KhanejaPRA2001}
\bibinfo{author}{\bibfnamefont{N.}~\bibnamefont{Khaneja}},
  \bibinfo{author}{\bibfnamefont{R.}~\bibnamefont{Brockett}}, \bibnamefont{and}
  \bibinfo{author}{\bibfnamefont{S.~J.} \bibnamefont{Glaser}},
  \bibinfo{journal}{Phys. Rev. A} \textbf{\bibinfo{volume}{63}},
  \bibinfo{pages}{032308} (\bibinfo{year}{2001}).

\bibitem[{\citenamefont{Kraus and Cirac}(2001)}]{KrausPRA01}
\bibinfo{author}{\bibfnamefont{B.}~\bibnamefont{Kraus}} \bibnamefont{and}
  \bibinfo{author}{\bibfnamefont{J.~I.} \bibnamefont{Cirac}},
  \bibinfo{journal}{Phys. Rev. A} \textbf{\bibinfo{volume}{63}},
  \bibinfo{pages}{062309} (\bibinfo{year}{2001}).

\bibitem[{\citenamefont{Zhang et~al.}(2003)\citenamefont{Zhang, Vala, Sastry,
  and Whaley}}]{ZhangPRA2003}
\bibinfo{author}{\bibfnamefont{J.}~\bibnamefont{Zhang}},
  \bibinfo{author}{\bibfnamefont{J.}~\bibnamefont{Vala}},
  \bibinfo{author}{\bibfnamefont{S.}~\bibnamefont{Sastry}}, \bibnamefont{and}
  \bibinfo{author}{\bibfnamefont{K.~B.} \bibnamefont{Whaley}},
  \bibinfo{journal}{\href{https://link.aps.org/doi/10.1103/PhysRevA.67.042313}{Phys.
  Rev. A}} \textbf{\bibinfo{volume}{67}}, \bibinfo{pages}{042313}
  (\bibinfo{year}{2003}).

\bibitem[{\citenamefont{Watts et~al.}(2013)\citenamefont{Watts, O'Connor, and
  Vala}}]{WattsE2013}
\bibinfo{author}{\bibfnamefont{P.}~\bibnamefont{Watts}},
  \bibinfo{author}{\bibfnamefont{M.}~\bibnamefont{O'Connor}}, \bibnamefont{and}
  \bibinfo{author}{\bibfnamefont{J.}~\bibnamefont{Vala}},
  \bibinfo{journal}{Entropy} \textbf{\bibinfo{volume}{15}},
  \bibinfo{pages}{1963} (\bibinfo{year}{2013}).

\bibitem[{\citenamefont{Watts et~al.}(2015)\citenamefont{Watts, Vala, M\"uller,
  Calarco, Whaley, Reich, Goerz, and Koch}}]{WattsPRA2015}
\bibinfo{author}{\bibfnamefont{P.}~\bibnamefont{Watts}},
  \bibinfo{author}{\bibfnamefont{J.~c.~v.} \bibnamefont{Vala}},
  \bibinfo{author}{\bibfnamefont{M.~M.} \bibnamefont{M\"uller}},
  \bibinfo{author}{\bibfnamefont{T.}~\bibnamefont{Calarco}},
  \bibinfo{author}{\bibfnamefont{K.~B.} \bibnamefont{Whaley}},
  \bibinfo{author}{\bibfnamefont{D.~M.} \bibnamefont{Reich}},
  \bibinfo{author}{\bibfnamefont{M.~H.} \bibnamefont{Goerz}}, \bibnamefont{and}
  \bibinfo{author}{\bibfnamefont{C.~P.} \bibnamefont{Koch}},
  \bibinfo{journal}{\href{https://link.aps.org/doi/10.1103/PhysRevA.91.062306}{Phys.
  Rev. A}} \textbf{\bibinfo{volume}{91}}, \bibinfo{pages}{062306}
  (\bibinfo{year}{2015}).

\bibitem[{\citenamefont{Goerz et~al.}(2015{\natexlab{a}})\citenamefont{Goerz,
  Gualdi, Reich, Koch, Motzoi, Whaley, Vala, M\"uller, Montangero, and
  Calarco}}]{GoerzPRA2015}
\bibinfo{author}{\bibfnamefont{M.~H.} \bibnamefont{Goerz}},
  \bibinfo{author}{\bibfnamefont{G.}~\bibnamefont{Gualdi}},
  \bibinfo{author}{\bibfnamefont{D.~M.} \bibnamefont{Reich}},
  \bibinfo{author}{\bibfnamefont{C.~P.} \bibnamefont{Koch}},
  \bibinfo{author}{\bibfnamefont{F.}~\bibnamefont{Motzoi}},
  \bibinfo{author}{\bibfnamefont{K.~B.} \bibnamefont{Whaley}},
  \bibinfo{author}{\bibfnamefont{J.}~\bibnamefont{Vala}},
  \bibinfo{author}{\bibfnamefont{M.~M.} \bibnamefont{M\"uller}},
  \bibinfo{author}{\bibfnamefont{S.}~\bibnamefont{Montangero}},
  \bibnamefont{and} \bibinfo{author}{\bibfnamefont{T.}~\bibnamefont{Calarco}},
  \bibinfo{journal}{Phys. Rev. A} \textbf{\bibinfo{volume}{91}},
  \bibinfo{pages}{062307} (\bibinfo{year}{2015}{\natexlab{a}}).

\bibitem[{\citenamefont{Makhlin}(2002)}]{MakhlinQIP2002}
\bibinfo{author}{\bibfnamefont{Y.}~\bibnamefont{Makhlin}},
  \bibinfo{journal}{Quantum Inf. Process.} \textbf{\bibinfo{volume}{1}},
  \bibinfo{pages}{243} (\bibinfo{year}{2002}).

\bibitem[{\citenamefont{Goerz et~al.}(2015{\natexlab{b}})\citenamefont{Goerz,
  Whaley, and Koch}}]{EPJQT.2.21}
\bibinfo{author}{\bibfnamefont{M.~H.} \bibnamefont{Goerz}},
  \bibinfo{author}{\bibfnamefont{K.~B.} \bibnamefont{Whaley}},
  \bibnamefont{and} \bibinfo{author}{\bibfnamefont{C.~P.} \bibnamefont{Koch}},
  \bibinfo{journal}{\href{https://doi.org/10.1140/epjqt/s40507-015-0034-0}{EPJ
  Quantum Technology}} \textbf{\bibinfo{volume}{2}}, \bibinfo{pages}{21}
  (\bibinfo{year}{2015}{\natexlab{b}}).

\bibitem[{\citenamefont{Jones et~al.}()\citenamefont{Jones, Oliphant, Peterson
  et~al.}}]{scipy}
\bibinfo{author}{\bibfnamefont{E.}~\bibnamefont{Jones}},
  \bibinfo{author}{\bibfnamefont{T.}~\bibnamefont{Oliphant}},
  \bibinfo{author}{\bibfnamefont{P.}~\bibnamefont{Peterson}},
  \bibnamefont{et~al.}, \emph{\bibinfo{title}{{SciPy}: Open source scientific
  tools for {Python}}}, \bibinfo{note}{\url{http://www.scipy.org/}}.

\bibitem[{\citenamefont{Byrd et~al.}(1995)\citenamefont{Byrd, Lu, Nocedal, and
  Zhu}}]{SIAMJSciComput.16.1190}
\bibinfo{author}{\bibfnamefont{R.~H.} \bibnamefont{Byrd}},
  \bibinfo{author}{\bibfnamefont{P.}~\bibnamefont{Lu}},
  \bibinfo{author}{\bibfnamefont{J.}~\bibnamefont{Nocedal}}, \bibnamefont{and}
  \bibinfo{author}{\bibfnamefont{C.}~\bibnamefont{Zhu}},
  \bibinfo{journal}{\href{http://dx.doi.org/10.1137/0916069}{SIAM J. Sci.
  Comput.}} \textbf{\bibinfo{volume}{16}}, \bibinfo{pages}{1190}
  (\bibinfo{year}{1995}).

\bibitem[{\citenamefont{Zhu et~al.}(1997)\citenamefont{Zhu, Byrd, Lu, and
  Nocedal}}]{ZhuATMS97}
\bibinfo{author}{\bibfnamefont{C.}~\bibnamefont{Zhu}},
  \bibinfo{author}{\bibfnamefont{R.~H.} \bibnamefont{Byrd}},
  \bibinfo{author}{\bibfnamefont{P.}~\bibnamefont{Lu}}, \bibnamefont{and}
  \bibinfo{author}{\bibfnamefont{J.}~\bibnamefont{Nocedal}},
  \bibinfo{journal}{ACM Trans. Math. Softw.} \textbf{\bibinfo{volume}{23}},
  \bibinfo{pages}{550} (\bibinfo{year}{1997}).

\bibitem[{\citenamefont{Nichol et~al.}(2017)\citenamefont{Nichol, Orona,
  Harvey, Fallahi, Gardner, Manfra, and Yacoby}}]{NPJQuantumInformation.3.3}
\bibinfo{author}{\bibfnamefont{J.~M.} \bibnamefont{Nichol}},
  \bibinfo{author}{\bibfnamefont{L.~A.} \bibnamefont{Orona}},
  \bibinfo{author}{\bibfnamefont{S.~P.} \bibnamefont{Harvey}},
  \bibinfo{author}{\bibfnamefont{S.}~\bibnamefont{Fallahi}},
  \bibinfo{author}{\bibfnamefont{G.~C.} \bibnamefont{Gardner}},
  \bibinfo{author}{\bibfnamefont{M.~J.} \bibnamefont{Manfra}},
  \bibnamefont{and} \bibinfo{author}{\bibfnamefont{A.}~\bibnamefont{Yacoby}},
  \bibinfo{journal}{\href{https://doi.org/10.1038/s41534-016-0003-1}{NPJ
  Quantum Information}} \textbf{\bibinfo{volume}{3}}, \bibinfo{pages}{3}
  (\bibinfo{year}{2017}).

\bibitem[{\citenamefont{Barends et~al.}(2014)\citenamefont{Barends, Kelly,
  Megrant, Veitia, Sank, Jeffrey, C., Mutus, Fowler, Campbell
  et~al.}}]{Nature.508.500503}
\bibinfo{author}{\bibfnamefont{R.}~\bibnamefont{Barends}},
  \bibinfo{author}{\bibfnamefont{J.}~\bibnamefont{Kelly}},
  \bibinfo{author}{\bibfnamefont{A.}~\bibnamefont{Megrant}},
  \bibinfo{author}{\bibfnamefont{A.}~\bibnamefont{Veitia}},
  \bibinfo{author}{\bibfnamefont{D.}~\bibnamefont{Sank}},
  \bibinfo{author}{\bibfnamefont{E.}~\bibnamefont{Jeffrey}},
  \bibinfo{author}{\bibfnamefont{W.~T.} \bibnamefont{C.}},
  \bibinfo{author}{\bibfnamefont{J.}~\bibnamefont{Mutus}},
  \bibinfo{author}{\bibfnamefont{A.~G.} \bibnamefont{Fowler}},
  \bibinfo{author}{\bibfnamefont{B.}~\bibnamefont{Campbell}},
  \bibnamefont{et~al.},
  \bibinfo{journal}{\href{http://dx.doi.org/10.1038/nature13171}{Nature}}
  \textbf{\bibinfo{volume}{508}}, \bibinfo{pages}{500503}
  (\bibinfo{year}{2014}).

\bibitem[{\citenamefont{{Wang} et~al.}(2012)\citenamefont{{Wang}, {Bishop},
  {Kestner}, {Barnes}, {Sun}, and {Das Sarma}}}]{WangNatComm2012}
\bibinfo{author}{\bibfnamefont{X.}~\bibnamefont{{Wang}}},
  \bibinfo{author}{\bibfnamefont{L.~S.} \bibnamefont{{Bishop}}},
  \bibinfo{author}{\bibfnamefont{J.~P.} \bibnamefont{{Kestner}}},
  \bibinfo{author}{\bibfnamefont{E.}~\bibnamefont{{Barnes}}},
  \bibinfo{author}{\bibfnamefont{K.}~\bibnamefont{{Sun}}}, \bibnamefont{and}
  \bibinfo{author}{\bibfnamefont{S.}~\bibnamefont{{Das Sarma}}},
  \bibinfo{journal}{Nature Communications} \textbf{\bibinfo{volume}{3}},
  \bibinfo{eid}{997} (\bibinfo{year}{2012}).

\bibitem[{\citenamefont{Kestner et~al.}(2013)\citenamefont{Kestner, Wang,
  Bishop, Barnes, and Das~Sarma}}]{KestnerPRL2013}
\bibinfo{author}{\bibfnamefont{J.~P.} \bibnamefont{Kestner}},
  \bibinfo{author}{\bibfnamefont{X.}~\bibnamefont{Wang}},
  \bibinfo{author}{\bibfnamefont{L.~S.} \bibnamefont{Bishop}},
  \bibinfo{author}{\bibfnamefont{E.}~\bibnamefont{Barnes}}, \bibnamefont{and}
  \bibinfo{author}{\bibfnamefont{S.}~\bibnamefont{Das~Sarma}},
  \bibinfo{journal}{Phys. Rev. Lett.} \textbf{\bibinfo{volume}{110}},
  \bibinfo{pages}{140502} (\bibinfo{year}{2013}).

\bibitem[{\citenamefont{Kogan}(1996)}]{ElectronicFluctuationsInSolids}
\bibinfo{author}{\bibfnamefont{S.}~\bibnamefont{Kogan}},
  \emph{\bibinfo{title}{Electronic Noise and Fluctuations in Solids}}
  (\bibinfo{publisher}{Cambridge University Press}, \bibinfo{year}{1996}).

\bibitem[{\citenamefont{Dial et~al.}(2013)\citenamefont{Dial, Shulman, Harvey,
  Bluhm, Umansky, and Yacoby}}]{PhysRevLett.110.146804}
\bibinfo{author}{\bibfnamefont{O.~E.} \bibnamefont{Dial}},
  \bibinfo{author}{\bibfnamefont{M.~D.} \bibnamefont{Shulman}},
  \bibinfo{author}{\bibfnamefont{S.~P.} \bibnamefont{Harvey}},
  \bibinfo{author}{\bibfnamefont{H.}~\bibnamefont{Bluhm}},
  \bibinfo{author}{\bibfnamefont{V.}~\bibnamefont{Umansky}}, \bibnamefont{and}
  \bibinfo{author}{\bibfnamefont{A.}~\bibnamefont{Yacoby}},
  \bibinfo{journal}{\href{https://link.aps.org/doi/10.1103/PhysRevLett.110.146804}{Phys.
  Rev. Lett.}} \textbf{\bibinfo{volume}{110}}, \bibinfo{pages}{146804}
  (\bibinfo{year}{2013}).

\end{thebibliography}

\end{document}